\documentstyle [12pt,eqsecnum,aps,amsfonts] {revtex}
\input epsf
\newcommand{\beq}{\begin{equation}}
\newcommand{\eeq}{\end{equation}}
\newcommand{\beqs}{\begin{eqnarray}}
\newcommand{\eeqs}{\end{eqnarray}}

\begin{document}

\baselineskip 6.0mm

\begin{center}

{\large\bf Chromatic Polynomials for $J(\prod H)I$ Strip Graphs and their 
Asymptotic Limits}

\vspace{4mm}

   Martin Ro\v{c}ek$^1$, Robert Shrock$^2$, and Shan-Ho Tsai$^3$

\vspace{6mm}

Institute for Theoretical Physics  \\ 
State University of New York       \\
Stony Brook, N. Y. 11794-3840

\vspace{4mm}

{\large\bf Abstract} 

\end{center} 
\small{
We calculate the chromatic polynomials $P$ for $n$-vertex strip graphs of the 
form $J(\prod_{\ell=1}^m H)I$, where $J$ and $I$ are various subgraphs on the 
left and right ends of the strip, whose bulk is comprised of $m$-fold 
repetitions of a subgraph $H$.  The strips have free boundary conditions in 
the longitudinal direction and free or periodic boundary conditions in the
transverse direction. This extends our earlier calculations for strip graphs 
of the form $(\prod_{\ell=1}^mH)I$.  We use a generating function method.  
From these results we compute the asymptotic limiting function 
$W=\lim_{n \to \infty}P^{1/n}$; for $q \in {\mathbb Z}_+$ this has physical 
significance as the ground state degeneracy per site (exponent of the ground 
state entropy) of the $q$-state Potts antiferromagnet on the given strip.  
In the complex $q$ plane, $W$ is an analytic function except on a certain 
continuous locus ${\cal B}$.  In contrast to the $(\prod_{\ell=1}^mH)I$ strip 
graphs, where ${\cal B}$ (i) is independent of $I$, and (ii) consists of arcs
and possible line segments that do not enclose any regions in the $q$ plane, 
we find that for some $J(\prod_{\ell=1}^m H)I$ strip graphs, ${\cal B}$ (i) 
does depend on $I$ and $J$, and (ii) can enclose regions in the $q$ 
plane. Our study elucidates the effects of different end subgraphs $I$ and 
$J$ and of boundary conditions on the infinite-length limit of the strip 
graphs. 

\vspace{4mm}

\begin{flushleft}

1 \  email: rocek@insti.physics.sunysb.edu \\
2 \  email: shrock@insti.physics.sunysb.edu \\
3 \  communicating author; email: tsai@insti.physics.sunysb.edu; address as of
May, 1998: Department of Physics and Astronomy, University of Georgia, 
Athens, GA 30602 \\

\end{flushleft}

\pagebreak

\newpage

\pagestyle{plain}
\pagenumbering{arabic}
\renewcommand{\thefootnote}{\arabic{footnote}}
\setcounter{footnote}{0}

\section{Introduction}

\normalsize

In this paper we generalize our previous study \cite{strip} of chromatic 
polynomials and their asymptotic limits on strip graphs.  The motivation for
this work is that there are substances that exhibit nonzero ground state
entropy, such as ice \cite{liebwu}.  A simple model that exhibits
ground state entropy is the $q$-state Potts antiferromagnet (PAF)
\cite{potts,wurev} for sufficiently large values of $q$.  The ground state
entropy is related here to the ground state degeneracy per site $W$ by
$S_0=k_B\ln W$. There is a direct connection with mathematical graph theory
via the elementary equality $Z(G,q,T=0)_{PAF}=P(G,q)$, where 
$Z(G,q,T=0)_{PAF}$ is the partition function of the zero-temperature
$q$-state Potts AF and $P(G,q)$ is the chromatic polynomial
\cite{birk,rtrev}, giving the number
of ways of coloring the $n$-vertex graph $G$ with $q$ colors such that no
adjacent vertices have the same color, and the consequent equality
\beq
W(\{G\},q) = \lim_{n \to \infty}P(G,q)^{1/n}
\label{w}
\eeq
where $\{G\}$ denotes the $n \to \infty$ limit of the family of graphs $G$
(e.g., a regular lattice, $\{G\} = \Lambda$ with some specified boundary
conditions). Since $P(G,q)$ is a polynomial, there is a natural generalization
of $q$ from $q \in {\mathbb Z}_+$ to $q \in {\mathbb R}_+$, and
indeed to $q \in {\mathbb C}$.
In general, $W(\{G\},q)$ is an analytic function in the $q$ plane
except along a certain continuous locus of points, which we denote ${\cal B}$
(and at possible isolated singularities which will not be important here).
In the limit as $n \to \infty$, the locus ${\cal B}$ forms by means of a
coalescence of a subset of the zeros of $P(G,q)$ (called chromatic zeros of
$G$) \cite{bds}-\cite{read91}. 

In a series of papers \cite{p3afhc}-\cite{wn} two of us have calculated and 
analyzed $W(\{G\},q)$, both for physical values of $q$ (via rigorous upper 
and lower bounds, large--$q$ series calculations, and Monte Carlo 
measurements) and for the generalization to complex values of $q$. In Ref. 
\cite{strip} we presented exact calculations of chromatic polynomials 
and the asymptotic limiting $W$ functions for strip graphs of a variety of 
regular lattices. 
For this purpose we developed and applied a generating function 
method. Specifically, we considered strip graphs of regular lattices of type 
$s$ of the form 
\beq
(G_s)_{m;I} = (\prod_{\ell=1}^m H)I
\label{stripi}
\eeq
where $s$ stands for square, triangular, honeycomb, etc.  Thus, 
such a graph is composed of $m$ repetitions of a subgraph $H$ 
attached to an initial subgraph $I$ on one end, which, by convention, we take 
to be the right end.  In the case of homogeneous strips, $I=H$. As is implicit
in 
the above notation, we picture such strip graphs as having a length of $L_x$ 
vertices in the horizontal ($x$) direction and a width of $L_y$ vertices in the
transverse (vertical, $y$) direction.  Since for the limit of infinite length
strips, the physical ground state degeneracy per site, $W$, and ground
state entropy, $S_0$, are independent of the boundary conditions used in the
longitudinal direction, we used free boundary conditions for technical 
simplicity in Ref. \cite{strip}.  (Homeomorphic expansions of these 
strip graphs were also studied in Ref. \cite{hs}.)

In the present work we generalize our previous study by considering strip 
graphs of regular lattices in which there may be special subgraphs on both 
the left and right ends of the strip: 
\beq
(G_s)_{m,J,I} = J(\prod_{\ell=1}^m H)I
\label{stripij}
\eeq
The end subsgraphs $I$ and $J$ may be different from each other and from 
the vertical rungs of the strip (thought of as a horizontal ladder).
As before, we use the simplest type of boundary
conditions -- free ones -- in the longitudinal direction, since in the limit 
of interest, namely that of infinite length, the physical ground state 
entropy is independent of these boundary conditions.  However, since the
size of the limit is finite in the transverse direction, it is of interest to
consider both free and periodic boundary conditions in this transverse 
direction, and we do that here. Again, we use the generating function method. 
For the infinite-length limit of $(\prod_{\ell=1}^mH)I$ strip graphs, we 
proved that the nonanalytic locus ${\cal B}$ is independent of $I$, which is
plausible, since in this limit, the subgraph $I$ occupies a vanishingly small 
part of the full strip graph \cite{strip}.  However, our study of the 
infinite-length limit of $J(\prod_{\ell=1}^mH)I$ strip graphs reveals that in
some cases, ${\cal B}$ does depend on the end subgraphs $I$ and $J$.  
Furthermore, whereas for
$(\prod_{\ell=1}^mH)I$ strip graphs we found that ${\cal B}$ consists of arcs 
and possible line segments that do not enclose any regions in the $q$ plane,
in contrast, for $J(\prod_{\ell=1}^m H)I$ strip graphs, we show here that in
certain cases ${\cal B}$ does enclose regions in the $q$ plane.  See also 
Ref. \cite{thesis} and a companion paper on strip graphs of a particular 
Archimedean lattice \cite{striparch} denoted as $(3^3 \cdot 4^2)$. 

It will be convenient to define the function 
\beq
D_k(q) = \frac{P(C_k,q)}{q(q-1)} = 
\sum_{s=0}^{k-2}(-1)^s {{k-1}\choose {s}} q^{k-2-s}
\label{dk}
\eeq
where $P(C_k,q)$ is the chromatic polynomial for the circuit
(cyclic) graph $C_k$ with $k$ vertices,
\beq
P(C_k,q) = (q-1)^k + (-1)^k(q-1)
\label{pck}
\eeq

This paper is organized as follows. In Section 2 we describe our present
implementation of our generating function method.  In sections 3-5 
we apply this method to strip graphs of the square, honeycomb and triangular
lattices.  In section 6 we study strips of a certain Archimedean lattice
known as the $(4 \cdot 8^2)$ lattice.  Sections 7 and 8 present results for
the square and triangular lattices with periodic boundary conditions in the
transverse direction (equivalently, these may be called cylindrical boundary
conditions).  Some further discussion and  conclusions are given in section 9. 

\vspace{12mm}

\section{Generating Function Method and End Subgraphs}

   We denote the chromatic polynomial describing the coloring of the strip
graph $(G_s)_{m,J,I}$ with $q$ colors as $P((G_s)_{m,J,I},q)$.  In the
generating function method, this chromatic polynomial is given by means of an
expansion, in terms of an auxiliary variable $x$, of a generating function
$\Gamma(G_{s,J,I},q,x)$: 
\beq
\Gamma(G_{s,J,I},q,x) = \sum_{m=0}^{\infty} P((G_s)_{m,J,I},q)x^m
\label{gamma}
\eeq
As before for strip graphs of the form (\ref{stripi}), we 
find that the generating function $\Gamma(G_{s,J,I},q,x)$ is a 
rational function of the form
\beq
\Gamma(G_{s,J,I},q,x) = \frac{{\cal N}(G_{s,J,I},q,x)}
{{\cal D}(G_{s,J,I},q,x)}
\label{gammagen}
\eeq
with (suppressing $J,I$ dependence in the notation)
\beq
{\cal N}(G_s,q,x) = \sum_{j=0}^{j_{max}} a_{G_s,j}(q) x^j
\label{n}
\eeq
and
\beq
{\cal D}(G_s,q,x) = 1 + \sum_{k=1}^{k_{max}} b_{G_s,k}(q) x^k
\label{d}
\eeq
where the $a_{G_s,i}$ and $b_{G_s,i}$ are polynomials in $q$. 
The degrees $j_{max}$ and $k_{max}$ of ${\cal N}(G_s,q,x)$ and 
${\cal D}(G_s,q,x)$
as polynomials in $x$ depend on the type and width of the strip $G_s$.

The method that we shall use for calculating the generating function is
based on the addition-contraction theorem from graph theory.  This was briefly
described in Ref. \cite{strip}, together with the equivalent
deletion-contraction theorem; here we shall give a more detailed discussion. 
We first recall the statement of the addition-contraction theorem:
let $G$ be a graph, and let $v$ and $v'$ be two non-adjacent vertices
in $G$.  Form (i) the graph $G_{add.}$ in which one adds a bond connecting $v$
and $v'$, and (ii) the graph $G_{contr.}$ in which one identifies $v$ and
$v'$.  Then the chromatic polynomial for $G$ is equal to the sum of the
chromatic polynomial for the graphs $G_{add.}$ and $G_{contr.}$.  For our
discussion, we may begin by taking $J$ and $I$ to be the same as the other
vertical rungs of the strip. By applying the 
addition-contraction theorem to the right-hand side of the initial $H$ 
subgraph in $(G_s)_m$, we obtain a set of strip graphs $(G'_{s,j})_m$ with 
complete \cite{complete} subgraphs labeled by $j$ on the right-hand end. 
For example, let us consider a
strip $(G_s)_m = \prod_{\ell=1}^m H$ where the width is such that to go from
one transverse end to the other one traverses at least four vertices 
(including the two vertices on the transverse boundaries), 
and label the four vertices on the right-hand longitudinal end, in sequence, 
as $v_1,v_2,v_3,v_4$. (For a strip of the square lattice with free transverse
boundary conditions, this example 
corresponds to a strip of width $L_y=4$. For other types of strip this may
correspond to different values of $L_y$, as illustrated by explicit 
calculations below.) Now apply the addition-contraction theorem to the
pair $v_1,v_4$; this yields the equation
\beq
P((G_s')_{m-1},q) = P((G'_{s,b(1,4)})_m,q) + P((G'_{s,v_1=v_4})_m,q)
\label{v1v4}
\eeq
where $(G'_{s,b(1,4)})_m$ denotes the strip graph with the right-hand end
modified by the addition of the bond connecting vertices $v_1$ and $v_4$, and
$(G'_{s,v_1=v_4})_m$ denotes the strip graph with the right-end modified by
identifying vertices $v_1$ and $v_4$.  Applying the theorem again to each of
these two strip graphs, one obtains the equation
\beqs
P((G'_s)_{m-1},q) & = & P((G'_{s,K_4})_m,q) + P((G'_{s,K_3(v_1=v_3)})_m,q) +
P((G'_{s,K_3(v_2=v_4)})_m,q) \nonumber \\
& & + P((G'_{s,K_2(v_1=v_3,v_2=v_4)})_m,q)+P((G'_{s,K_3(v_1=v_4)})_m,q)
\label{addconend}
\eeqs
Let us label the five resultant complete graphs on the right-hand end via the
label $i$, with chromatic polynomials denoted by $P((G'_{s,i})_0,q)$. Thus, 
\beq
P((G'_{s,1})_0,q)=P(K_4,q)=q(q-1)(q-2)(q-3)
\label{pk4}
\eeq
\beq
P((G'_{s,i})_0,q)=P(K_3,q)=q(q-1)(q-2) \quad {\rm for} \quad i=2,3,5
\label{pk3}
\eeq
\beq
P((G'_{s,4})_0,q)=P(K_2,q)=q(q-1)
\label{pk2}
\eeq
Next, working one's way leftward from the right-hand end,
apply the addition-contraction theorem in a similar manner to
the next set of transverse vertices, and label the five resultant complete
graphs with the label $j$. Define a (square) matrix $M$ (in this case
$5 \times 5$) with elements consisting of the chromatic polynomials
of this set of graphs with complete subgraphs $i$ on the right and $j$ on
the left. This procedure transforms the initial strip into a sum of factorized
strips with parts that overlap in complete graphs. Therefore the intersection
theorem yields, for $m \ge 1$, 
\beq
P((G'_{s,j})_m,q)=\sum_i\frac{M_{ji}P((G'_{s,i})_{m-1},q)}
{P((G'_{s,i})_0,q)}=\sum_i (MD)_{ji}P((G'_{s,i})_{m-1},q),
\label{inteq}
\eeq
where $D$ is a ($5 \times 5$) \underline diagonal matrix with elements
\beq
D_{i,i}=\frac{1}{P((G'_{s,i})_0,q)} \quad 1 \le i \le 5
\label{dii}
\eeq
with the ordering as given above. (Do not confuse this matrix $D$ with the 
denominator ${\cal D}$ of the generating function $\Gamma$ defined in eq. 
(\ref{d}). 
Because of (\ref{addconend}), the generating function of $G'_s$ can be
written as
\beq
\Gamma(G'_s,q,x) = \sum_j \Gamma(G'_{s,j},q,x)
\label{gammags}
\eeq
Writing the generating function of $G'_{s,j}$ in the form
\beq
\Gamma(G'_{s,j},q,x) = \sum_{m=1}^{\infty} P((G'_{s,j})_m,q)x^{m-1}
\label{gammagsj}
\eeq
and using (\ref{inteq}) with (\ref{gammagsj}), we obtain
\beqs
\Gamma(G'_{s,j},q,x) &=& P((G'_{s,j})_1,q) + \sum_{i} x \ (MD)_{ji}
\sum_{m=2}^{\infty} P((G'_{s,i})_{m-1},q)x^{m-2} \cr\cr
&= & P((G'_{s,j})_1,q) + \sum_{i} x \ (MD)_{ji} \Gamma(G'_{s,i},q,x)
\label{gammagsjsubs}
\eeqs
The special case $m=1$ in eq. (\ref{inteq}) yields 
\beq
P((G'_{s,j})_1,q) = \sum_i M_{ji}
\label{chrompjmeq1}
\eeq
Let us define a vector ${\bf \Gamma}$ with elements 
$\Gamma(G_{s,j}',q,x)$, where the index $j$ also labels the elements of the 
vector. Thus, eq. (\ref{gammags}) can be written as
\beq
\Gamma(G'_s,q,x) = {\bf v_b}^T \cdot {\bf \Gamma}
\label{gammagscdot}
\eeq
where ${\bf v_b}$ is a vector with all elements equal to unity. In the example
described above, ${\bf v_b} = (1,1,1,1,1)^T$. (Do not confuse the vectors 
${\bf v_b}$ with the vertices $v_i$.) 
Similarly, let us define the vector ${\bf P_1}$ with elements 
$P((G'_{s,j})_1,q)$. Then eq. (\ref{chrompjmeq1}) yields
\beq
{\bf P_1} = M {\bf v_a}
\label{chromp1vec}
\eeq
where in this case ${\bf v_a}={\bf v_b}$. 
In vector form, eq. (\ref{gammagsjsubs}) becomes
\beq
{\bf \Gamma} = {\bf P_1} + x M D {\bf \Gamma}
\label{gpxmd}
\eeq
Solving for ${\bf \Gamma}$, we get
\beq
{\bf \Gamma} = (1-xMD)^{-1} {\bf P_1}
\label{gammavec}
\eeq
Henceforth, for notational simplicity, we drop the prime on $G$ and indicate
the end subgraphs $I$ and $J$ explicitly. 
Using (\ref{gammagscdot}) and (\ref{chromp1vec}) in eq. (\ref{gammavec}), we
obtain
\beq
\Gamma(G_{s,J,I},q,x)={\bf v_b}^T (1-x MD)^{-1} M {\bf v_a}={\bf v_b}^T
\sum_{m=0}^{\infty} x^m (MD)^m M {\bf v_a} ,
\label{gammaaddcon}
\eeq
where in this case, for the strip $G_s$, ${\bf v_a}=(1,1,1,1,1)^T$ and 
${\bf v_b}={\bf v_a}$. 
Note that in general ${\bf v_a}$ and ${\bf v_b}$ can be chosen 
independently, and different choices of these vectors correspond to different 
end graphs $J$ and $I$, respectively.  Thus, we can equivalently write 
(\ref{gamma}) as 
\beq
P((G_{s,J,I})_m,q)={\bf v_b}^T (MD)^m M {\bf v_a}
\label{addconpol}
\eeq
Factorizing the denominator of the generating function (\ref{gammaaddcon}), we
have 
\beq
{\cal D}(G_{s,J,I},q,x)=\prod_r (1-\lambda_r(q) x),
\label{addconden}
\eeq
where the 
${\lambda_r(q)}'s$ are eigenvalues of the product of matrices $MD$. The
subgraphs on the two ends of the strip (defined by ${\bf v_a}$ and
${\bf v_b}$) determine which eigenvalues enter in the product in equation
(\ref{addconden}).  Using this method, we have calculated chromatic
polynomials and their asymptotic limits for various strip graphs of the form
$(G_{s,J,I})_m = J(\prod_{\ell=1}^m H)I$.  We define $(G_{s,J,I})_{m=0}$ as
being equal to $JI$, where the end graphs $I$ and $J$ are, in general, 
different.  For technical convenience, we also use a slightly different
labelling convention than in Ref. \cite{strip}, which can be illustrated as
follows: for strips of the square lattice, for example, we take the end 
graphs $I$ and $J$ to refer to the respective sets of vertical bonds on the
right and left ends of the strip, whereas in Ref. \cite{strip}, $I$ referred 
to the first vertical layer of squares.  We now present 
some of these calculations and the analytic structure of the corresponding 
$W(\{G_{sJ,I}\},q)$.  

\vspace{6mm}

We start by defining the vectors which will be used to describe the subgraphs 
on the ends of the strips. Let us denote the four 
vertices on one end, in sequence, as $v_i$, with $i=1,...,4$ and 
consider end subgraphs to be described by vectors ${\bf v_a}$ and 
${\bf v_b}$  chosen from a set of possibilities that we list below: 

\vspace{3mm} 

(i) 
\beq
{\bf v_1}=(1,0,0,0,0)^T
\label{v1}
\eeq
which corresponds to a strip with a complete graph
$K_4$ on the 
end, as shown in Fig. \ref{stripbc8}(a) for a strip of the square lattice of 
width $L_y=4$. This $K_4$ is formed by adding edges between vertices $v_1$ and
$v_4$, $v_1$ and $v_3$, and $v_2$ and $v_3$. 

\vspace{2mm}

(ii)

\beq
{\bf v_2}=(0,1,0,0,0)^T
\label{v2}
\eeq
corresponding to a strip with vertices $v_1$ and
$v_3$ contracted and an extra edge connecting vertices $v_2$ and $v_4$,
yielding a $K_3(v_1=v_3)$ on the end. Fig. \ref{stripbc8}(b)
illustrates this for a strip of the square lattice of width $L_y=4$.

\vspace{2mm}

(iii) 

\beq
{\bf v_3}=(0,0,1,0,0)^T
\label{v3}
\eeq
which corresponds to a strip with vertices 
$v_2$ and $v_4$ contracted and an extra edge connecting vertices $v_1$ and 
$v_3$, yielding a $K_3(v_2=v_4)$ on the end. Fig. 
\ref{stripbc8}(c) illustrates this for a strip of the 
square lattice of width $L_y=4$.

\vspace{2mm}

(iv) 

\beq
{\bf v_4}=(0,0,0,1,0)^T
\label{v4}
\eeq
which corresponds to a strip with vertices 
$v_1$ contracted with $v_3$ and $v_2$ contracted with $v_4$, forming a 
$K_2(v_1=v_3,v_2=v_4)$ on the end. Fig. \ref{stripbc8}(d) illustrates this 
for a strip of the square lattice of width $L_y=4$.

\vspace{2mm}

(v)

\beq
{\bf v_5}=(0,0,0,0,1)^T
\label{v5}
\eeq
which corresponds to a strip with vertices 
$v_1$ and $v_4$ contracted, yielding a $K_3(v_1=v_4)$ on the end. Fig. 
\ref{stripbc8}(e) illustrates this for a strip of the 
square lattice of width $L_y=4$.

\vspace{2mm}

(vi)

\beq
{\bf v_6}=(1,1,0,0,0)^T
\label{v6}
\eeq
which corresponds to a strip with one extra
edge connecting vertices $v_1$ and $v_4$ and another connecting vertices $v_2$
and $v_4$. Fig. \ref{stripbc8}(f) illustrates this for a 
strip of the square lattice of width $L_y=4$.

\vspace{2mm}

(vii)

\beq
{\bf v_7}=(1,0,1,0,0)^T
\label{v7}
\eeq
which corresponds to a strip with one extra
edge connecting vertices $v_1$ and $v_4$ and another connecting vertices $v_1$
and $v_3$. Fig. \ref{stripbc8}(g) illustrates this for a 
strip of the square lattice of width $L_y=4$. Note that for a strip of the
square lattice this end subgraph is equivalent to that 
described by vector ${\bf v_6}=(1,1,0,0,0)^T$. However, for other types of 
strips ${\bf v_7}$ may represent a different end subgraph. 

\vspace{2mm}

(viii)

\beq
{\bf v}=(1,1,1,1,1)^T
\label{v}
\eeq
where the end subgraph is the same as the vertical rungs of the strip. 
Fig. \ref{stripbc8}(h) illustrates this 
for a strip of the square lattice of width $L_y=4$.
Fig. \ref{stripbc8} illustrates these end subgraphs for 
a strip of the square lattice of width $L_y=4$. The different subgraphs 
are shown on the right end of the strip. Similar subgraphs can be 
considered on the left end of the strip. In the remainder of this section we 
consider the end subgraphs of the strips described by the set of vectors above,
eqs. (\ref{v1})-(\ref{v}). 

\begin{figure}
\centering
\leavevmode
\epsfxsize=4.5in
\epsffile{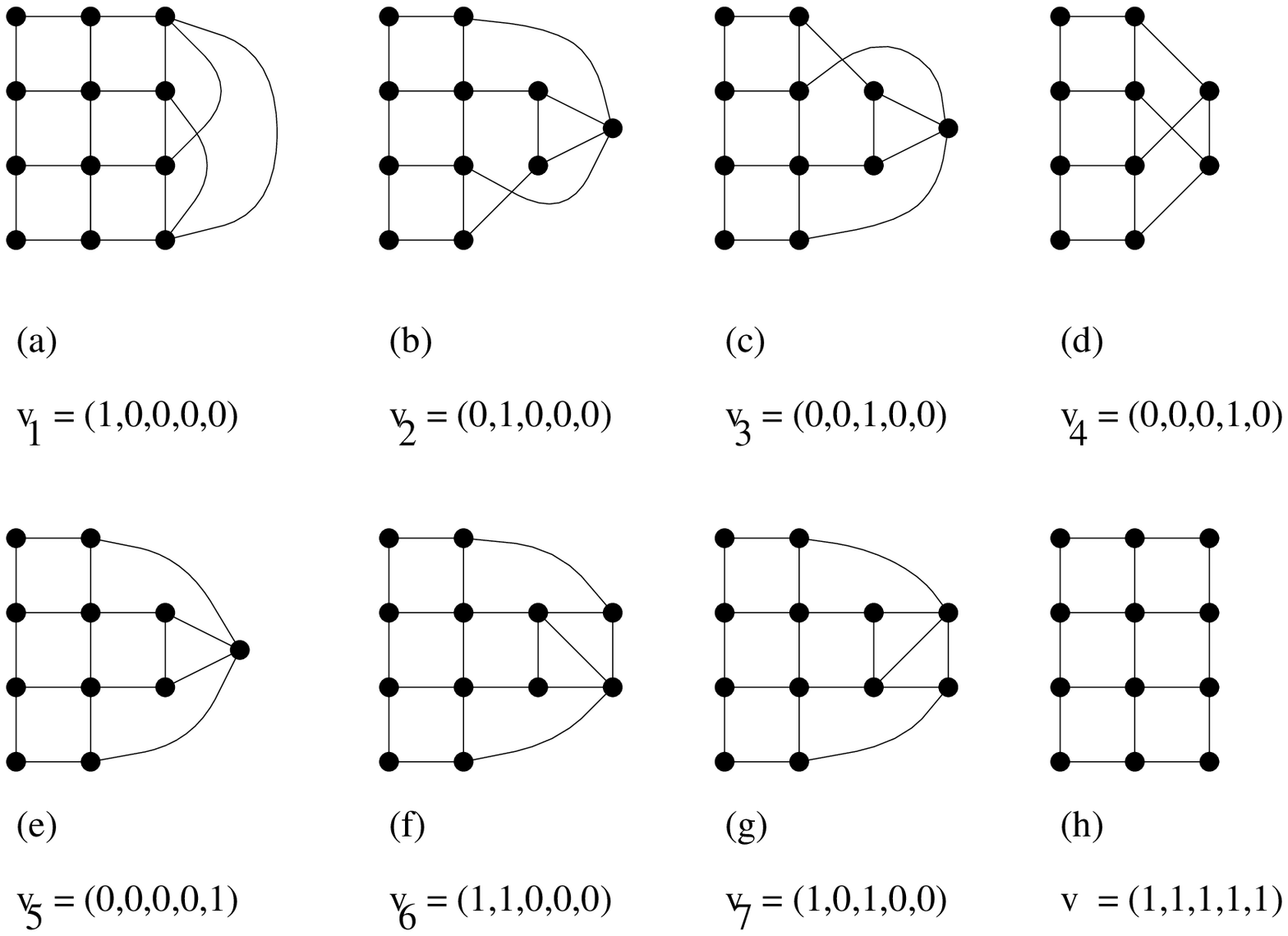}
\caption{\footnotesize{Illustration of end subgraphs on a strip of the
square lattice of width $L_y=4$ and free transverse boundary conditions.
Cases (a), (b), (c) and (d) contain
non-planar end subgraphs, while cases (e), (f), (g) and (h) contain planar 
end subgraphs.}}
\label{stripbc8}
\end{figure}

We label free and periodic boundary conditions in the
transverse ($y$) direction as
\beq
FBC_y \quad , \quad PBC_y
\label{bc}
\eeq
respectively, and similarly for the longitudinal ($x$) direction. The strips
have FBC$_x$ and, unless otherwise stated, also $FBC_y$. 
For strip graphs with FBC$_x$ and FBC$_y$, 
we shall restrict ourselves to cases where the product 
of matrices $MD$ is a $5\times 5$ matrix. 
For strip graphs with FBC$_x$ and PBC$_y$ 
we shall restrict ourselves to cases where the product 
of matrices $MD$ is a $4\times 4$ matrix. These are strip
graphs with cylindrical boundary conditions, with transverse cross sections
comprised by four vertices and four edges, forming squares. 

Let us denote the four vertices 
belonging to the transverse cross section of the strip on one of the 
longitudinal ends of the strip by $v_i$, with $i=1,...,4$, in sequence. The 
product of matrices $MD$ is a $4\times 4$ matrix in this case and vectors 
describing end subgraphs can be

(i) 
\beq
{\bf v_1}=(1,0,0,0)^T
\label{v1L4}
\eeq
which corresponds to a strip with a $K_4$ on the 
end. This $K_4$ is formed by adding edges connecting vertices $v_1$ and $v_3$
and vertices $v_2$ and $v_4$.

\vspace{2mm}

(ii) 
\beq
{\bf v_2}=(0,1,0,0)^T
\label{v2L4}
\eeq
corresponding to a strip with vertices $v_1$ and
$v_3$ contracted and an extra edge connecting vertices $v_2$ and $v_4$,
yielding a $K_3(v_1=v_3)$ on the end.

\vspace{2mm}

(iii) 
\beq
{\bf v_3}=(0,0,1,0)^T
\label{v3L4}
\eeq
which corresponds to a strip with vertices $v_2$ 
and $v_4$ contracted and an extra edge connecting vertices $v_1$ and $v_3$,
yielding a $K_3(v_2=v_4)$ on the end.

\vspace{2mm}

(iv) 
\beq
{\bf v_4}=(0,0,0,1)^T
\label{v4L4}
\eeq
which corresponds to a strip with vertices $v_2$ 
and $v_4$ contracted and vertices $v_1$ and $v_3$ contracted forming a
$K_2(v_1=v_3,v_2=v_4)$ on the end.

\vspace{2mm}

(v) 
\beq
{\bf v_5}=(1,1,0,0)^T
\label{v5L4}
\eeq
where the ending subgraph is equal to the repeating
unit with an extra edge connecting the non-adjacent pair of vertices $v_2$ and
$v_4$.

\vspace{2mm}

(vi) 
\beq
{\bf v_6}=(1,0,1,0)^T
\label{v6L4}
\eeq
similar to (v), with the extra edge connecting
the other pair of non-adjacent vertices, namely, $v_1$ and $v_3$.

\vspace{2mm}

(vii) 
\beq
{\bf v}=(1,1,1,1)^T
\label{vL4}
\eeq
where the ending subgraph is equal to the vertical rungs along the strip. 

In our studies of strip graphs with PBC$_y$ we only
consider subgraphs on the two free ends of the strips described
by the vectors listed in eqs. (\ref{v1L4})-(\ref{vL4}).
We next present results for a number of strip graphs of type $G_s'$. 
 
\section{Strip of the Square Lattice of Width $L_{\lowercase{y}}=4$}

For a strip of the square ($sq$) lattice of width $L_y=4$ vertices and 
FBC$_y$, we have (with $f_{sq,i} \equiv f_i$ to save space) 
\beq
M =  \left (\begin{array}{ccccc}
(q-3)f_{0}f_{6} \quad &(q-3)f_{0}f_{1}\quad &(q-3)f_{0}f_{1} &(q-3)f_{0}f_{2} 
&(q-3)f_{0}f_{1} \\
(q-3)f_{0}f_{1} \quad & f_{0}f_{3} \quad & (q-2)f_{0}f_{2} & f_{0}f_{5} 
&f_{0}f_{4} \\
(q-3)f_{0}f_{1} \quad & (q-2)f_{0}f_{2}\quad & f_{0}f_{3} & f_{0}f_{5}
&f_{0}f_{4} \\
(q-3)f_{0}f_{2} \quad &  f_{0}f_{5}\quad & f_{0}f_{5} & q(q-1)D_4 &
(q^2-5q+7)f_{0} \\
(q-3)f_{0}f_{1} \quad & f_{0}f_{4}\quad & f_{0}f_{4} & (q^2-5q+7)f_{0}
& f_{0}f_{3}
\end{array}\right )
\label{mmatrixsq}
\eeq
where
\beq
f_{sq,0}=q(q-1)(q-2)
\label{matrixelemsqf0}
\eeq
\beq
f_{sq,1}=q^3-7q^2+19q-20
\label{matrixelemsqf1}
\eeq
\beq
f_{sq,2}=q^2-5q+8
\label{matrixelemsqf2}
\eeq
\beq
f_{sq,3}=q^3-6q^2+14q-13
\label{matrixelemsqf3}
\eeq
\beq
f_{sq,4}=q^3-7q^2+18q-17
\label{matrixelemsqf4}
\eeq
\beq
f_{sq,5}=q^2-4q+5
\label{matrixelemsqf5}
\eeq
\beq
f_{sq,6}=q^4-10q^3+41q^2-84q+73
\label{matrixelemsqf6}
\eeq
Hence, using eq. (\ref{dii}) with eqs. (\ref{pk4})-(\ref{pk2}), we obtain the 
matrix product $MD$:
\beq
MD =  \left (\begin{array}{ccccc}
f_{6} \qquad & (q-3)f_{1}\qquad & (q-3)f_{1} & (q-2)(q-3)f_{2} &
(q-3)f_{1} \\
f_{1} \qquad & f_{3} \qquad& (q-2)f_{2} & (q-2)f_{5} & f_{4} \\
f_{1} \qquad & (q-2)f_{2}\qquad & f_{3} & (q-2)f_{5} & f_{4} \\
f_{2} \qquad & f_{5}\qquad & f_{5} & D_4 & q^2-5q+7 \\
f_{1} \qquad & f_{4}\qquad & f_{4} & (q-2)(q^2-5q+7) & f_{3}
\end{array}\right )
\label{mdmatrixsq}
\eeq

Three of the eigenvalues of this $MD$ matrix, which we label as $\lambda_i$, 
$i=1,2,3$ are the inverses of the three roots of the 
polynomial 
\beq
1+b_{sq(4),1}x+b_{sq(4),2}x^2+b_{sq(4),3}x^3
\label{dsq}
\eeq
where \cite{strip} 
\beq
b_{sq(4),1} = -q^4+7q^3-23q^2+41q-33
\label{b1sqw3}
\eeq
\beq
b_{sq(4),2} = 2q^6-23q^5+116q^4-329q^3+553q^2-517q+207
\label{b2sqw3}
\eeq
\beq
b_{sq(4),3} = -q^8+16q^7-112q^6+449q^5-1130q^4+1829q^3-1858q^2+1084q-279
\label{b3sqw3}
\eeq
For all end subgraphs studied here, namely, considering any two pairs
among (\ref{v1})-(\ref{v}) 
as the subgraphs on the two ends of the strip, these three
eigenvalues contribute to the generating functions of the respective strips.
The other two eigenvalues are $\lambda_4=1$ and $\lambda_5=(q^2-4q+3)$. 
Depending on the end subgraphs chosen, 
either $\lambda_4$ or $\lambda_5$ or both contribute to the denominator of the
generating function for the chromatic polynomial. However, we find that 
$\lambda_4$ and $\lambda_5$ are never leading eigenvalues.  Hence, the 
nonanalytic locus ${\cal B}$ of $W(\{G_{sq(4),J,I}\},q)$ does not depend on the
subgraphs $J$ and $I$ on the two ends of the strip and is given by Fig. 3(b) 
of Ref. \cite{strip}.  Of course, the actual chromatic polynomials and their
zeros for specific finite strips do depend on $J$ and $I$.

\section{Strip of the Honeycomb Lattice of Width $L_{\lowercase{y}}=3$}

For a strip of the honeycomb ($hc$) lattice of width $L_y=3$ and 
FBC$_y$, we compute the $MD$ matrix in the same manner as was discussed in
detail above for the square lattice.  We find that 
three of the eigenvalues of the $MD$ matrix are the inverses of the roots of 
\beq
1+b_{hc(3),1}x+b_{hc(3),2}x^2+ b_{hc(3),3}x^3
\label{dhc}
\eeq
where the coefficients $b_{hc(3),1}$, $b_{hc(3),2}$ and 
$b_{hc(3),3}$ are given by eqs. (3.14)--(3.16) in Ref. \cite{strip}; to
save space, we do not list them here. 
For all end subgraphs studied here, namely, considering 
any two pairs among (\ref{v1})-(\ref{v}) as the subgraphs on the two ends 
of the strip, these three eigenvalues contribute to the generating functions 
of the respective strips. The other two eigenvalues are identically zero. 
Therefore, just as was the case with the square lattice of width $L_y=4$, the 
denominator of the generating function, and thus ${\cal B}$, are
not modified by the subgraphs on the two ends of the strip.

\section{Strip of the Triangular Lattice of Width $L_{\lowercase{y}}=4$}

For a strip of the triangular ($t$) lattice with width $L_y=4$, we find that 
four of these five eigenvalues of the $MD$ matrix 
are the inverses of the roots of the polynomial 
\beq
1+b_{t(4),1}x+b_{t(4),2}x^2+b_{t(4),3}x^3+b_{t(4),4}x^4
\label{dtrily4}
\eeq
where the 
coefficients $b_{t(4),1}$, $b_{t(4),2}$, $b_{t(4),3}$ and $b_{t(4),4}$ were 
given respectively by eqs. (B.15)--(B.18) in Ref. \cite{strip}.  Since their
expressions are somewhat lengthy, we do not list them here. 
Let us refer to these four eigenvalues as $\lambda_i$, with $i=1,2,3,4$.
For all end subgraphs studied here, namely, considering any two pairs
among (\ref{v1})-(\ref{v}) 
as the subgraphs on the two ends of the strip, these four
eigenvalues contribute to the generating functions of the respective strips.
The fifth eigenvalue is $\lambda_5=1$, which contributes to the generating
function for certain end subgraphs. $\lambda_5$ is leading in a region
of the complex $q$ plane and thus modify the curves ${\cal B}$ of 
non-analyticities of $W(\{G_{t(4),J,I}\},q)$. In Table \ref{tabletribc} we 
show for various planar and non-planar end subgraphs $J$ and $I$, 
whether or not $\lambda_5$ 
contributes to the generating function, and some features of the boundary
${\cal B}$ for each case.  In turn, the presence or absence of $\lambda_5$ in
the generating function is determined by the end subgraphs $I$ and $J$, or
equivalently, the vectors ${\bf v}_a$ and ${\bf v}_b$ via the basic equation 
(\ref{addconpol}).  Indeed, one sees from Table \ref{tabletribc} that 
${\cal B}$ encloses regions if and only if $\lambda_5$ appears in the
generating function.  This phenomenon, in which ${\cal B}$ can depends on 
the end subgraphs of the strip, is reminiscent of the 
dependence of certain aspects of complex-temperature phase boundaries and 
partition function zeros of the Potts model on boundary conditions found in 
Refs. \cite{martin,pfef}.  

\pagebreak 

\begin{table}
\caption{\footnotesize{
Illustrative end subgraphs and properties of generating function
for a strip of the triangular lattice of 
width $L_y=4$. The notation $J$,$I$ means that the vectors ${\bf v_j}$ and
${\bf v_i}$ describe the end subgraphs on the two ends of the strip. P stands
for planar and NP for non-planar. The third column shows whether or
not $\lambda_5$ enters in the generating function. The fourth column lists
some features of the boundary ${\cal B}$. }}
\begin{center}
\footnotesize
\begin{tabular}{|cccc|} \hline
$J,I$ & planarity & does $\lambda_5$ enter? & features of ${\cal B}$\\
\hline\hline
${\bf v},{\bf v}$     & P,P & N & arcs \\
${\bf v},{\bf v_i}$, $i=1,2,3,4$ & P,NP   & N & arcs \\
${\bf v},{\bf v_i}$, $i=5,6,7$   & P,P & N & arcs \\
${\bf v_i},{\bf v_i}$, $i=1,2,3,4$    & NP,NP & Y & arcs and one 
enclosed region\\
${\bf v_i},{\bf v_i}$, $i=5,6,7$    & P,P & N & arcs \\
${\bf v_1},{\bf v_i}$, $i=2,3,4$ & NP,NP & Y & arcs and one enclosed region\\
${\bf v_1},{\bf v_i}$, $i=5,6,7$ & NP,P & N & arcs\\
${\bf v_6},{\bf v_i}$, $i=1,2,3,4$ & P,NP & N & arcs\\
${\bf v_6},{\bf v_i}$, $i=5,7$ & P,P & N & arcs\\
\hline
\end{tabular}
\normalsize
\end{center}
\label{tabletribc}
\end{table}
 
In Fig. \ref{striptrizw3m10bc}(a) we show the analytic structure of the $W$ 
function for an infinitely long strip of the triangular lattice with width 
$L_y=4$ and end subgraphs $J,I$ that do not have $\lambda_5$ 
in the generating function.  The resultant nonanalytic locus ${\cal B}$ is the
same as that which we found in Ref. \cite{strip} (see Fig. 5(b) of that paper)
for the infinitely long strip of the triangular lattice of the form 
$(\prod_{\ell=1}^{\infty} H)I$ with width $L_y=4$, for arbitrary $I$.
The locus ${\cal B}$ in Fig. \ref{striptrizw3m10bc}(a) consists of arcs 
that do not separate the complex $q$ plane into separate regions.  This locus
includes complex-conjugate multiple points of valence three; these are singular
points on the algebraic curve constituted by ${\cal B}$ in the terminology of
algebraic geometry \cite{alg}.  Similarly, multiple points are observed in many
of the loci ${\cal B}$ for strip graphs to be described below. 
Since the locus ${\cal B}$ forms by coalescence of zeros of the chromatic 
polynomial (called chromatic zeros of the graph) as the length of the strip 
graph goes to infinity, it is of
interest to compare the locations of these zeros for a reasonably long finite
strip with the asymptotic locus ${\cal B}$.  Accordingly, we show in this 
figure the chromatic zeros for a strip with $I=H$ and length $m=12$.

In Fig. \ref{striptrizw3m10bc}(b) we display the analytic 
structure of the $W$ function
for a strip of the triangular lattice with end subgraphs that yield 
$\lambda_5$ in the generating function. In this case the locus ${\cal B}$ is
comprised by arcs which do enclose a self-conjugate region
where $\lambda_5$ is the leading eigenvalue. The boundaries of this region
cross the real $q$ axis at $q=q_1 = \frac{3+\sqrt{5}}{2}=2.618...$ 
and at $q = 3$.  Following our earlier work \cite{p3afhc,w}, we define $q_c$ 
as the maximal finite real  
value of $q$ where $W$ is nonanalytic for a given lattice, i.e.,
where ${\cal B}$ crosses the real axis.  Hence, for the present strip, 
$q_c=3$. We note that for the infinite triangular lattice, $q_c=4$ 
\cite{w}.  For comparison, in \ref{striptrizw3m10bc}(b) we exhibit 
the zeros of the chromatic polynomial 
$P(({G_{t(4),J,I}})_m,q)$ with subgraphs on the right and left
ends of the strip described by $J=I={\bf v}_2$ and $m=10$ repeating units in 
the strip.

We can understand this difference in the
topology of the loci ${\cal B}$ as follows.  One complex-conjugate 
pair of arc endpoints in the 
locus ${\cal B}$ in Fig. \ref{striptrizw3m10bc}(a) occurs at 
$q \simeq  2.759\pm 0.154i$.  However, for the triangular $L_y=4$ strips with 
endgraphs that yield $\lambda_5$ in the generating function, this pair of
would-be endpoints, and the contiguous portion of the associated arcs, do not
actually occur on ${\cal B}$ because they lie inside the enclosed region in 
Fig. \ref{striptrizw3m10bc}(b) where $\lambda_5$ is leading.  Thus, as was 
anticipated in our discussion in Ref. \cite{strip}, the difference in topology
of ${\cal B}$ for the $(\prod_{\ell=1}^{\infty} H)I$ strips and the present 
$J(\prod_{\ell=1}^{\infty} H)I$ strip of the triangular lattice is associated 
with the
fact that for the former types of strips, the leading $\lambda$'s were always
algebraic, whereas here a leading $\lambda$, viz., $\lambda_5=1$ is a
(constant) polynomial, rather than an algebraic function of $q$.  Here we see
how these features that ${\cal B}$ can (i) depend on $J,I$ and (ii) enclose
regions are both connected with each other via the basic eq. 
(\ref{addconpol}), which determines which $\lambda$'s appear in the generating
function and, in particular, whether the leading $\lambda$'s are always
algebraic, as for $(\prod_{\ell=1}^{\infty} H)I$ strips with nontrivial 
${\cal B}$ loci, or can also include polynomial $\lambda$'s, as for 
$J(\prod_{\ell=1}^{\infty} H)I$ strips. 

\pagebreak

\begin{figure}
\vspace{-4cm}
\centering
\leavevmode
\epsfxsize=3.0in
\begin{center}
\leavevmode
\epsffile{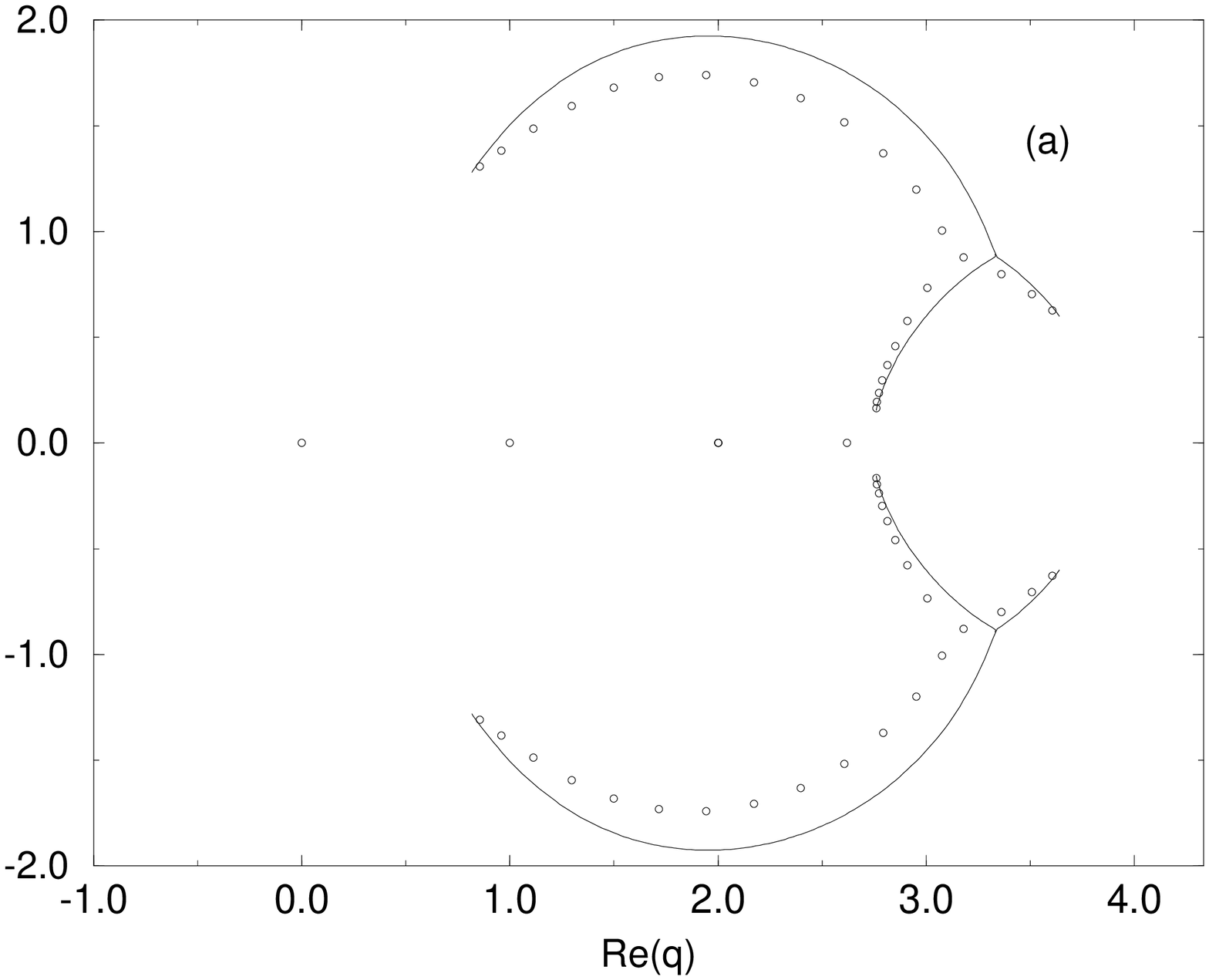}
\end{center}
\vspace{-3cm}
\begin{center}
\leavevmode
\epsfxsize=3.0in
\epsffile{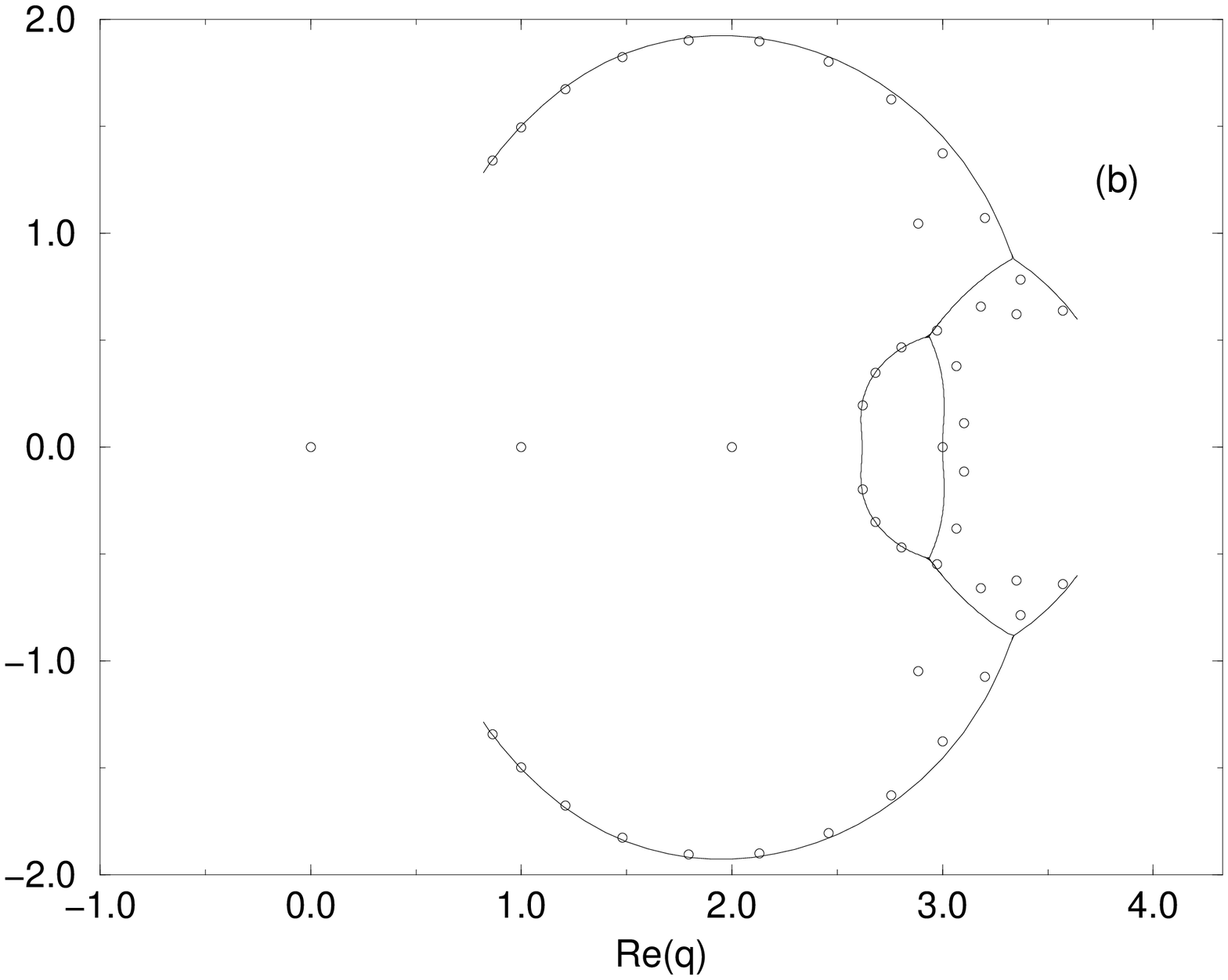}
\end{center}
\vspace{-2cm}
\caption{\footnotesize{Analytic structure of the function 
$W(\{G_{t(4),J,I}\},q)$ for a (a) $(L_x=\infty) \times (L_y=4)$ strip of the
triangular lattice given by $\lim_{m \to \infty} 
J(\prod_{\ell=1}^m H)I$ with end graphs $I,J$ that do not yield $\lambda_5$ in
the generating function (this includes the case where $J$ is absent); 
(b) $\lim_{m \to \infty} J(\prod_{\ell=1}^m H)I$ with end subgraphs $I,J$ 
that yield $\lambda_5$ in the generating function.  For comparison, the 
zeros of the chromatic polynomials are shown for the strips with (a) $I=H$ 
and $m=12$ (whence $n=56$); (b) subgraphs $I=J$ described by ${\bf v_2}$ and 
$m=10$ (whence $n=46$).}}
\label{striptrizw3m10bc}
\end{figure}

\vspace{2mm}

Thus, with changes in the subgraphs on the two ends $I$ and $J$ 
of a strip of the 
triangular lattice of width $L_y=4$, the analytic structure of 
$W(\{G_{t(4),J,I}\},q)$ changes from a set of arcs to structures which enclose
regions. For this type of strip we only obtained enclosed regions when the
subgraphs on both ends of the strip were non-planar. However, in Ref. 
\cite{striparch} an example is given where this happens with planar end
subgraphs, albeit with a heteropolygonal Archimedean lattice, viz., the 
$(3^3 \cdot 4^2)$ lattice. 
Our present calculation shows that for the strip graphs of 
the form $J(\prod_{\ell=1}^\infty H)I$, the nonanalytic locus
${\cal B}$ may depend on the end subgraphs $I$ and $J$, in contrast to the 
property proved in Ref. \cite{strip} that for strip graphs of the 
form $(\prod_{\ell=1}^\infty H)I$, the resultant ${\cal B}$ is independent 
of the end subgraph $I$. 

\vspace{3 mm}

\section{Strip of the $(4 \cdot 8^2)$ Lattice of Width $L_{\lowercase{y}}=3$}

We have also studied strips of lattices that are
tilings involving more than one type of regular polygon.  We recall
\cite{gs,cmo,wn,strip} that an Archimedean lattice is a uniform tiling of the 
plane by one or more regular polygons, such that each vertex is equivalent to 
every other vertex.  An Archimedean lattice is identified by the symbol 
\beq
\Lambda = (\prod_i p_i^{a_i})
\label{archlambda}
\eeq 
where the product is the ordered sequence of polygons that one traverses
in making a complete circuit around a vertex in a given (say counterclockwise)
direction, and the notation $p_i^{a_i}$ indicates that the regular $p$-sided
polygon $p_i$ occurs contiguously $a_i$ times in this circuit.  In this section
we consider a strip of the $(4 \cdot 8^2)$ lattice of width $L_y=3$ with free
boundary conditions in the $y$ direction.  For this strip, the $MD$ matrix is $
5 \times 5$, and three of the eigenvalues of this matrix are the inverses of
the roots of $1+b_{488(3),1}x+b_{488(3),2}x^2+ b_{488(3),3}x^3$, where the
coefficients $b_{488(3),j}$, $j=1,2,3$ were given in Ref. \cite{strip}.  For
all of the end graphs $I$ and $J$ studied here, namely considering any two
pairs among (i)-(viii) in eqs. (\ref{v1})--(\ref{v}), these three eigenvalues
contribute to the generating functions of the respective strips.  The other two
eigenvalues are identically zero.  Therefore, as was the case with the strip of
the honeycomb lattice with $L_y=3$, the denominator of the generating function
and thus ${\cal B}$, are not modified by the end graphs $I$ and $J$ on the
strip.

\section{Strip of the Square Lattice with PBC$_{\lowercase{y}}$}

In this section we consider a strip of square lattice with PBC$_y$, i.e., 
periodic boundary conditions in the transverse direction (= cylindrical 
boundary conditions). 
We begin by observing that if one takes a strip of a regular lattice with 
FBC$_y$ and identifies the vertices with lowest and highest 
values of $y$ for each $x$ 
to produce PBC$_y$, the changes in ${\cal B}$ depend on 
the specific type of lattice and value of $L_y$.  For
example, let us consider the square lattice with FBC$_y$ and $L_y=4$ and
carry out the above-mentioned operation, equivalent to gluing together the
upper and lower edges of the strip, so that the transverse cross sections are
triangles.   Although the infinite-length limit of the
original strip with FBC$_y$ yielded a nontrivial ${\cal B}$, shown in Fig. 
3(b) of Ref. \cite{strip}, after the gluing operation, the chromatic
polynomial factorizes trivially for the square strip with PBC$_y$ and 
triangular cross section, and ${\cal B}=\emptyset$, i.e. there is no
locus of points where $W$ is nonanalytic in the $q$ plane.  In contrast, for a
number of cases to be discussed below, the gluing operation takes an infinite
strip with nontrivial ${\cal B}$ to another infinite strip with nontrivial and
different ${\cal B}$.  As we shall show below, for strips of the square 
(triangular) lattice with transverse cross sections forming squares, and end
graphs $I$ and $J$ that are the same as the other vertical rungs of the 
strip, the resultant ${\cal B}$ does not (does) enclose regions, respectively. 

We begin by studying the strip of the square lattice with PBC$_y$ and
transverse cross sections forming squares.  
Depending on one's labelling conventions, this corresponds to 
$L_y=4$ or $L_y=5$, where in the latter case, one interprets the periodic
boundary conditions as identifying the top and bottom vertices for each value
of $x$.  It is interesting to observe that this graph can be 
regarded as part of a three dimensional simple cubic lattice. For end 
subgraphs $I$ and $J$ described by the vector ${\bf v}$ 
we obtain a generating function with $j_{max}=1$, $k_{max}=2,$ and coefficients
\beq
a_{sq(4),c,0}=q(q-1)(q^6-11q^5+55q^4-159q^3+282q^2-290q+133)
\label{a0sqtb}
\eeq
\beq
a_{sq(4),c,1}=-q(q-1)(q^2-3q+3)(q^6-12q^5+61q^4-169q^3+269q^2-231q+85)
\label{a1sqtb}
\eeq
\beq
b_{sq(4),c,1}=-q^4+8q^3-29q^2+55q-46
\label{b1sqtb}
\eeq
\beq
b_{sq(4),c,2}=q^6-12q^5+61q^4-169q^3+269q^2-231q+85
\label{b2sqtb}
\eeq
In this case the $MD$ matrix is the upper left-hand $4 \times 4$ submatrix of
$MD$ given above in eq. (\ref{mdmatrixsq}) for the square strip with 
FBC$_y$. Two of the four eigenvalues of the 
$MD$ matrix are the inverses of the roots
of $1+b_{sq(4),c,1} x+b_{sq(4),c,2} x^2$. These two eigenvalues, denoted
$\lambda_1$ and $\lambda_2$,
contribute to the generating functions for all end graphs $J$ and $I$ that we
consider.  The other two eigenvalues are
$\lambda_3=1$ and $\lambda_4= (q-1)(q-3)$, which contribute to the generating
functions for some end graphs. In Table \ref{tablesqtb} we show for
various end subgraphs $J,I$ whether 
or not $\lambda_3$ and  $\lambda_4$ contribute to the generating function and
some features of the boundaries ${\cal B}$ for each case.

\begin{table}
\caption{\footnotesize{
Illustrative end subgraphs and properties of generating function for a 
strip of the square lattice with 
periodic boundary conditions on the transverse direction and transverse
cross sections forming squares.  Notation is as in Table \ref{tabletribc}. 
The second and third columns indicate whether or not $\lambda_3$ and
$\lambda_4$, respectively, contribute to the generating function.}}
\begin{center}
\footnotesize
\begin{tabular}{|cccc|} \hline
$J,I$ & does $\lambda_3$ enter? & does $\lambda_4$ enter? & features of
${\
\cal B}$\\
\hline\hline
${\bf v},{\bf v}$    & N & N & arcs \\
${\bf v_i},{\bf v_i}$, $i=1,4$ & Y & N & arcs \\
${\bf v_i},{\bf v_i}$, $i=2,3$ & Y &Y& arcs and one pair of enclosed regions\\
${\bf v_i}/{\bf v_i}$, $i=5,6$ & N &Y& arcs and one pair of enclosed regions\\
${\bf v},{\bf v_i}$, $i=1,..,6$ & N & N & arcs \\
${\bf v_1},{\bf v_i}$, $i=2,3,4$ & Y & N & arcs \\
${\bf v_1},{\bf v_i}$, $i=5,6$ & N & N & arcs \\
${\bf v_2},{\bf v_3}$ & Y & Y & arcs and one pair of enclosed regions\\
${\bf v_2},{\bf v_i}$, $i=5,6$ & N &Y & arcs and one pair of enclosed regions\\
${\bf v_3},{\bf v_i}$, $i=5,6$ & N &Y & arcs and one pair of enclosed regions\\
${\bf v_4},{\bf v_i}$, $i=2,3$ & Y &N & arcs \\
${\bf v_4},{\bf v_i}$, $i=5,6$ & N &N & arcs \\
${\bf v_5},{\bf v_6}$ & N & Y & arcs and one pair of enclosed regions\\
\hline
\end{tabular}
\normalsize
\end{center}
\label{tablesqtb}
\end{table}

\begin{figure}
\centering
\leavevmode
\epsfxsize=3.5in
\begin{center}
\leavevmode
\epsfxsize=3.5in
\epsffile{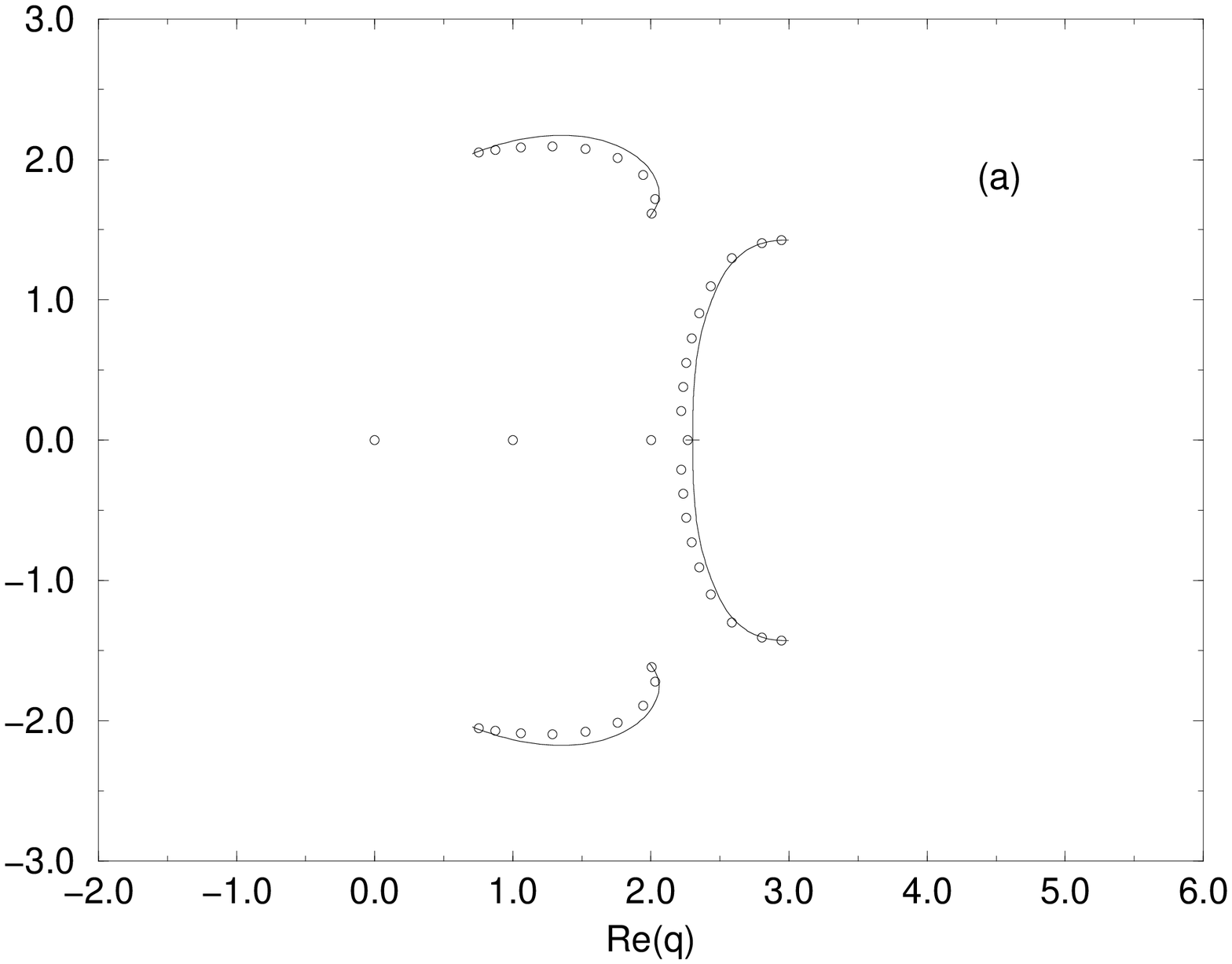}
\end{center}
\vspace{-4cm}
\begin{center}
\leavevmode
\epsfxsize=3.5in
\epsffile{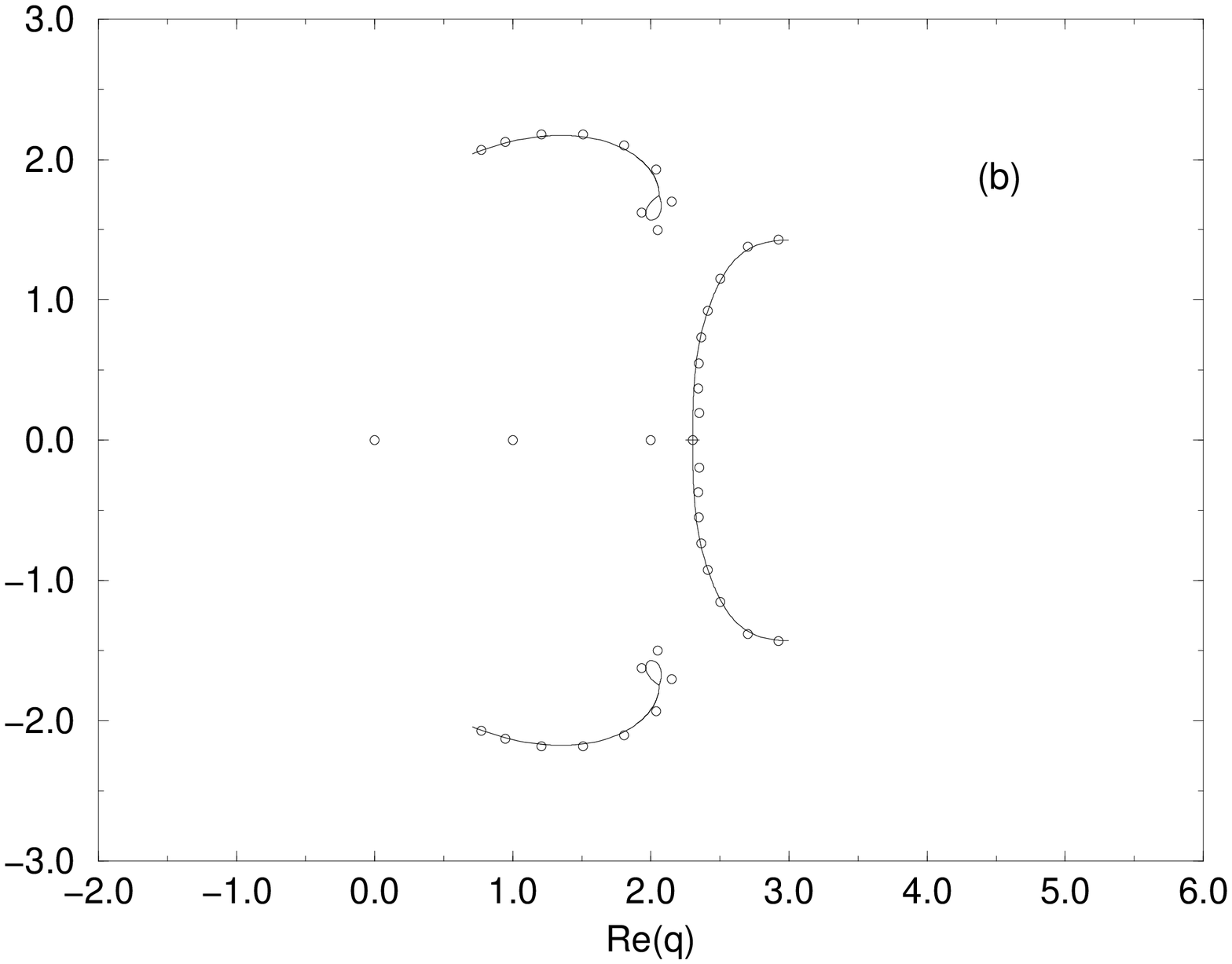}
\end{center}
\vspace{-2cm}
\caption{\footnotesize{Analytic structure of the function $W$ for an 
infinitely long strip 
of square lattice with periodic boundary condition on the transverse
direction and transverse cross sections forming squares: 
(a) corresponds to end graphs $I$ and $J$ for which
$\lambda_4$ does not enter in the generating function. (b) corresponds to
cases where $\lambda_4$ enters in the generating function.
For comparison, the zeros of the chromatic
polynomial for a strip with $m=8$ repeating units ($n=40$ vertices in (a) and
$n=36$ vertices in (b)) and end graphs $I$ and $J$ described by (a) ${\bf v}$
[(b) ${\bf v_2}$] on the right and left ends of the strip, are shown. }}
\label{figsqtb}
\end{figure}

We note that although $\lambda_3$ enters in the generating functions for
some end graphs, it is never leading, so the analytic structure
is not modified by its presence in the generating function. For end graphs 
where $\lambda_4$ is not present in the generating function
the nonanalytic locus ${\cal B}$ of the $W$ function, shown in Fig. 
\ref{figsqtb}(a), includes one complex-conjugate
pair of arcs and one self-conjugate arc which crossing the real $q$ axis at
$q\simeq 2.3026$. There is also a line segment on the real $q$ axis from
$q= 2.2534$ to $q=2.3517$, which latter point is thus the value of $q_c$ for 
(the $m \to \infty$ limit of) this strip.  The locus ${\cal B}$ in Fig. 3(a) 
is somewhat similar to that which we presented in Fig. 3(a) of Ref. 
\cite{strip} for the infinitely long strip of the 
square lattice of width $L_y=3$ with FBC$_y$ 
except that in that case, the self-conjugate arc crossed the real $q$
axis at $q=2$, which was thus the value of $q_c$ in that case.  The value of 
$q_c$ that we find for PBC$_y$, viz., $q_c \simeq 2.35$, is comparable to the 
value $q_c \simeq 2.27$ that we found for the
infinitely long strip of the square lattice with FBC$_y$ 
(see Fig. 3(b) of Ref. \cite{strip}) and width $L_y=4$. 
The endpoints of the arcs and the line segment
correspond to the branch points of the eigenvalues $\lambda_1$ and
$\lambda_2$.
For end graphs $I$ and $J$  where $\lambda_4$ is present in the generating
function the analytic structure of the $W$ function, shown in Fig.
\ref{figsqtb}(b), has an extra pair of complex-conjugate regions. In these
enclosed regions the eigenvalue $\lambda_4$ is leading. The complex-conjugate
pair of branch points of the eigenvalues $\lambda_1$ and $\lambda_2$ at
$q=1.992\pm 1.594i$ lie within the complex-conjugate enclosed regions, where
$\lambda_4$ is leading. Hence, these branch points do not correspond to arc
endpoints.

\section{Strip of the Triangular Lattice with PBC$_{\lowercase{y}}$}

We next consider a strip of the triangular lattice with PBC$_y$ 
and transverse cross sections forming 
squares.  As above, depending on one's labelling conventions, this corresponds
to $L_y=4$ or $L_y=5$, where in the latter case one identifies the top and
bottom vertices for each value of $x$.  The subgraphs on the horizontal ends of
the strip are described by the vector ${\bf v}$.  We obtain 
a generating function with $j_{max}=1$, $k_{max}=2,$ and coefficients
\beq
a_{t(4),c,0}=q(q-1)(q-2)(q-3)(q^4-10q^3+41q^2-84q+71)
\label{a0ttb}
\eeq
\beq
a_{t(4),c,1}=-2q(q-1)(q-2)(q-3)^3(q^2-5q+5)(q^2-3q+3)
\label{a1ttb}
\eeq
\beq
b_{t(4),c,1}=-(q-3)(q^3-9q^2+33q-48)
\label{b1ttb}
\eeq
\beq
b_{t(4),c,2}=2(q-2)(q-3)^3(q^2-5q+5)
\label{b2ttb}
\eeq
Two of the four eigenvalues of the $MD$ matrix are the inverses of the roots
of $1+b_{t(4),c,1} x+b_{t(4),c,2} x^2$. These two eigenvalues, denoted
$\lambda_1$ and $\lambda_2$ for the $+$ and $-$ signs of the square root,
respectively, contribute to the generating functions for all 
end graphs $I$ and $J$ that we consider. The other two eigenvalues are
$\lambda_3=0$ and $\lambda_4=2$, the latter contributes to the generating
functions for some end graphs. In Table \ref{tablettb} we show for
various subgraphs $J,I$ on the two ends of the strips, whether or not
$\lambda_4$ contributes to the generating function and some
features of the boundary ${\cal B}$ for each case.

\pagebreak

\begin{table}
\caption{\footnotesize{
Illustrative end subgraphs and properties of generating function for a 
strip of the 
triangular lattice with periodic boundary conditions on the transverse 
direction and transverse cross sections forming squares. The notation is as in 
Table \ref{tabletribc}.  The second column indicates whether or
not $\lambda_4$ contributes to the generating function.}} 

\begin{center}
\footnotesize
\begin{tabular}{|ccc|} \hline
boundaries & does $\lambda_4$ enter? & features of ${\cal B}$\\
\hline\hline
${\bf v}/{\bf v}$   & N & arcs and closed region (a)\\
${\bf v}/{\bf v_i}$, $i=1,..,6$   & N & arcs and closed region (a)\\
${\bf v_1}/{\bf v_i}$, $i=1,..,6$ & Y & arcs and closed region (b)\\
${\bf v_2}/{\bf v_i}$, $i=2,..,6$ & Y & arcs and closed region (b)\\
${\bf v_3}/{\bf v_i}$, $i=3,..,6$ & Y & arcs and closed region (b)\\
${\bf v_4}/{\bf v_i}$, $i=4,..,6$ & Y & arcs and closed region (b)\\
${\bf v_5}/{\bf v_i}$, $i=5,6$   & N & arcs and closed region (a)\\
${\bf v_6}/{\bf v_6}$   & N & arcs and closed region (a)\\
\hline
\end{tabular}
\normalsize
\end{center}
\label{tablettb}
\end{table}

\begin{figure}
\centering
\leavevmode
\epsfxsize=3.5in
\begin{center}
\leavevmode
\epsfxsize=3.5in
\epsffile{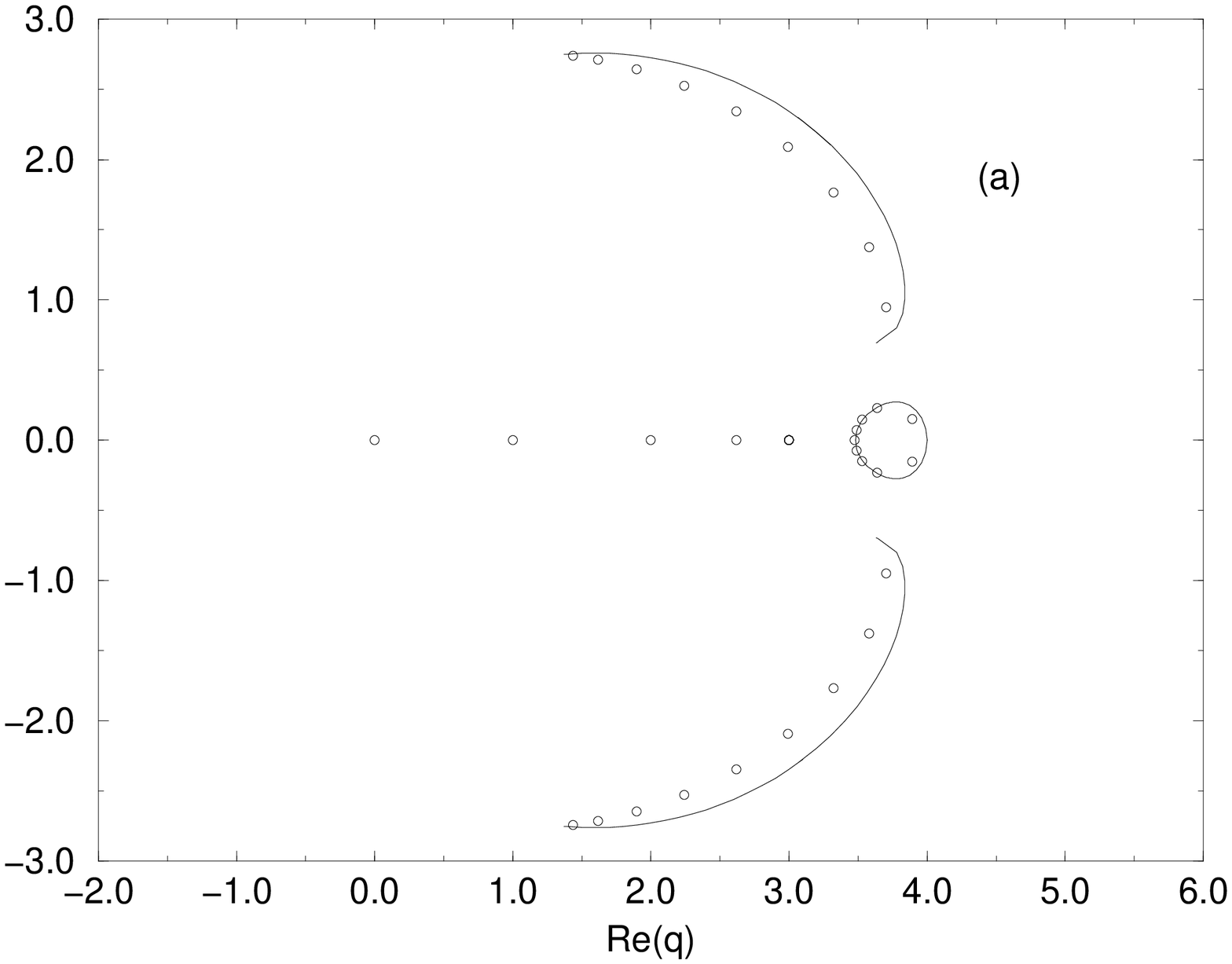}
\end{center}
\vspace{-4cm}
\begin{center}
\leavevmode
\epsfxsize=3.5in
\epsffile{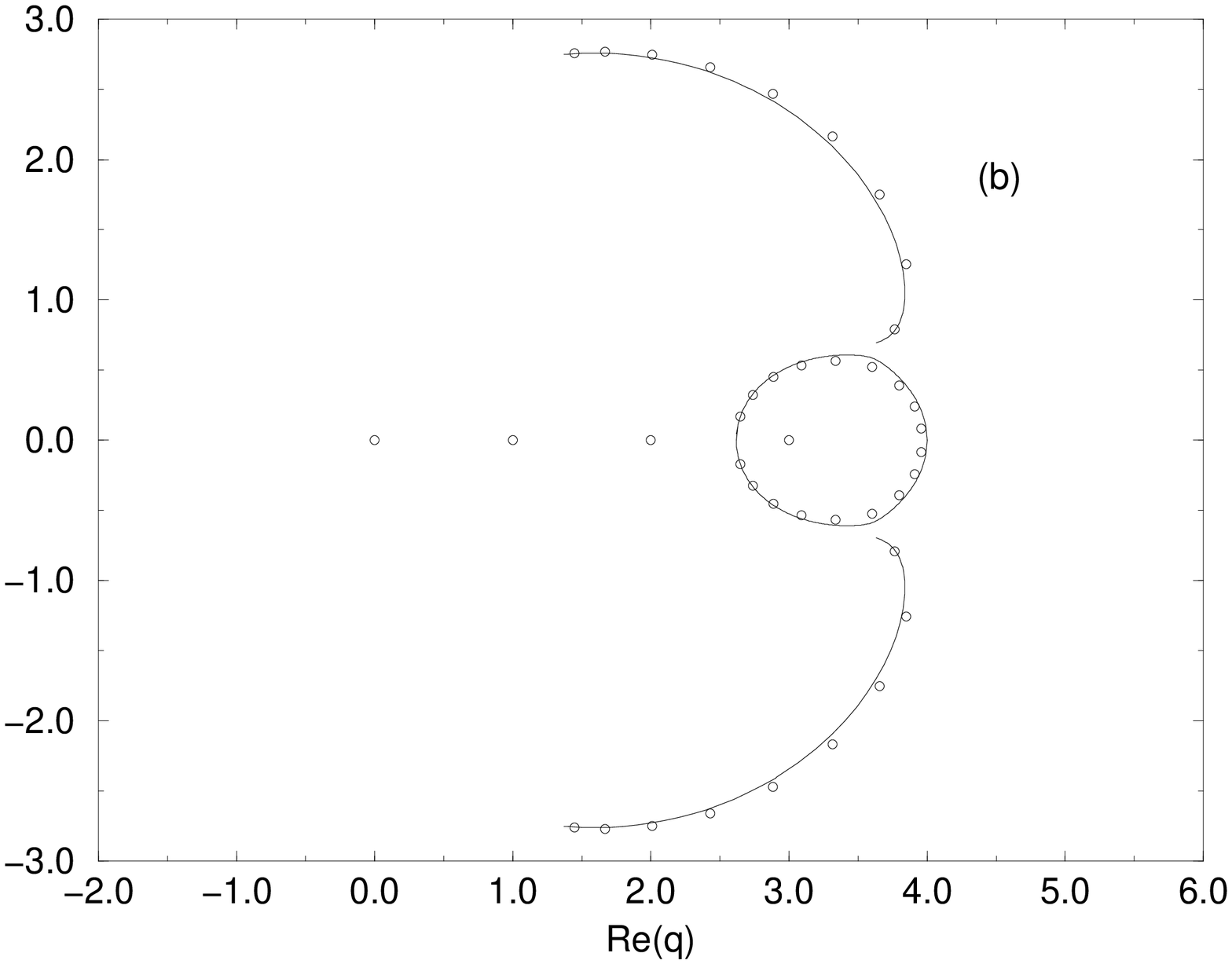}
\end{center}
\vspace{-2cm}
\caption{\footnotesize{Analytic structure of the function $W$ for the strip
of triangular lattice with periodic boundary condition on the transverse
direction, transverse cross sections forming squares, and length $L_x=\infty$.
(a) corresponds to end graphs $I$ and $J$ for which 
$\lambda_4$ does not enter in the generating function. (b) corresponds to
cases where $\lambda_4$ enters in the generating function.
For comparison, the zeros of the chromatic
polynomial for a strip with $m=8$ repeating units ($n=40$ vertices) and
end graphs $I$ and $J$ described by (a) ${\bf v}$
[(b) ${\bf v_1}$] on the right and left ends of the strip, are shown. }}
\label{figttb}
\end{figure}

For end graphs where $\lambda_4$ is not present in the generating
function the analytic structure of the $W$ function, shown in
Fig. \ref{figttb}(a), is formed by one pair of complex-conjugate arcs
and one self-conjugate region. Outside (inside) the enclosed region 
$\lambda_1$ ($\lambda_2$) is leading,  The endpoints of the arcs correspond
to the branch points of the eigenvalues $\lambda_1$ and $\lambda_2$. The
boundary ${\cal B}$ crosses the real $q$ axis at $q\simeq 3.481$ and at
$q=4$. For end graphs where $\lambda_4$ is present in the generating
function the analytic structure of the $W$ function, shown in
Fig. \ref{figttb}(b), is qualitatively the same as the previous case,
except that the enclosed region is larger and $\lambda_4$ is leading therein.
The boundary ${\cal B}$ crosses the real $q$ axis at
$q=\frac{3+\sqrt{5}}{2}=2.618...$ and at $q=4$. It is interesting to note that
for a strip of the triangular lattice with PBC$_y$ and 
transverse cross sections forming squares and different end graphs $I$ and $J$,
our results yield a value of $q_c$ which coincides with the value for an
infinite triangular lattice, namely, $q_c=4$ \cite{w}.  However,
this is not a general result; in our analysis above of the strip of
the square lattice with PBC$_y$ and transverse cross sections forming squares,
we obtained $q_c \simeq 2.35$, which is somewhat less than the value 
$q_c=3$ for the infinite square lattice.

\section{Discussion and Conclusions} 

In this paper we have presented exact calculations of chromatic polynomials 
$P$ and asymptotic limiting $W$ functions for strip graphs of the form 
$J(\prod_{\ell=1}^\infty H)I$, where $J$ and $I$ are 
various subgraphs on the left and right ends of the strip, and 
$(\prod_{\ell=1}^m H)$ are strips of various regular lattices consisting of 
$m$-fold repetitions of subgraph units $H$.  We have also studied the effects
of using periodic as well as free boundary conditions in the transverse
direction.  For the respective strip graphs we 
have determined the loci ${\cal B}$ where $W$ is 
nonanalytic in the complex $q$ plane.  
Our present analysis generalizes our earlier calculations for 
strip graphs of the form $(\prod_{j=1}^\infty H)I$.  

In contrast to the $(\prod_{\ell=1}^\infty H)I$ strip
graphs, where ${\cal B}$ (i) is independent of $I$, and (ii) consists of arcs
and possible line segments that do not enclose any regions in the $q$ plane,
we find that for some $J(\prod_{\ell=1}^\infty H)I$ strip graphs, 
${\cal B}$ (i) does depend on $I$ and $J$, and (ii) can enclose regions in 
the $q$ plane.  We have explained these differences in the present work and
have related them, via eq. (\ref{addconpol}) to whether the leading $\lambda$'s
are algebraic, as in the case of $(\prod_{\ell=1}^\infty H)I$ strips with
nontrivial ${\cal B}$, or can also include polynomial $\lambda$'s, as in
certain of the $J(\prod_{\ell=1}^{\infty} H)I$ strips.  It should be noted that
even if all of the leading $\lambda$'s are algebraic, the resultant locus 
${\cal B}$ can still enclose regions.  This can occur when this locus ${\cal
B}$ is noncompact in the $q$ plane, extending infinitely far from the origin
\cite{wa}.  Indeed, we showed in Ref. \cite{wa} that whenever ${\cal B}$ is
noncompact in the $q$ plane, passing through $1/q=0$, it encloses regions.  An
example of a family of graphs with only algebraic $\lambda$'s which has a locus
${\cal B}$ that encloses regions (and is noncompact) is the $\{U_k\}$ family
discussed in detail in (the third paper of) Ref. \cite{wa}.  Our present study
thus elucidates the effects of different end subgraphs $I$ and $J$ and of free
versus periodic transverse boundary conditions on the function
$W$ describing the ground state degeneracy, per site, of the Potts
antiferromagnet on infinitely long 2D strips of various lattices. 

\vspace{10mm} 

\begin{center}

{\large\bf Acknowledgments} 

\end{center}

This research was supported in part by the NSF grant PHY-97-22101.

\pagebreak

\vfill
\eject

\begin{thebibliography}{99}
\bibitem{complete}{The complete graph $K_n$ on $n$ vertices is defined as the
graph where each of these vertices is connected to all of the other $n-1$
vertices by bonds.  The complement $\overline K_n$ of $K_n$ is the graph with
$n$ vertices and no bonds.}
\end{thebibliography}
\end{document}